\documentclass[sigconf]{acmart}
\usepackage{balance}
\def\BibTeX{{\rm B\kern-.05em{\sc i\kern-.025em b}\kern-.08emT\kern-.1667em\lower.7ex\hbox{E}\kern-.125emX}}

 \setcopyright{none}
\renewcommand\footnotetextcopyrightpermission[1]{} 

\usepackage{multirow}

\copyrightyear{2019} 
\acmYear{2019} 
\acmConference[CIKM '19]{The 28th ACM International Conference on Information and Knowledge Management}{November 3--7, 2019}{Beijing, China}
\acmBooktitle{The 28th ACM International Conference on Information and Knowledge Management (CIKM '19), November 3--7, 2019, Beijing, China}
\acmPrice{15.00}
\acmDOI{10.1145/3357384.3358096}
\acmISBN{978-1-4503-6976-3/19/11}

\setcopyright{acmlicensed}

\begin{document}
\fancyhead{}
\title{Towards More Usable Dataset Search: From Query Characterization to Snippet Generation}

\author{Jinchi Chen}
\affiliation{%
  \institution{National Key Laboratory for Novel Software Technology, Nanjing University, China}
}
\email{jcchen@smail.nju.edu.cn}
\author{Xiaxia Wang}
\affiliation{%
  \institution{National Key Laboratory for Novel Software Technology, Nanjing University, China}
}
\email{xxwang@smail.nju.edu.cn}
\author{Gong Cheng}
\affiliation{%
  \institution{National Key Laboratory for Novel Software Technology, Nanjing University, China}
}
\email{gcheng@nju.edu.cn}
\author{Evgeny Kharlamov}
\affiliation{%
  \department{Bosch Center for AI}
  \institution{Bosch GmbH, Germany}
  \institution{University of Oslo, Norway}
}
\email{evgeny.kharlamov@de.bosch.com}
\email{evgeny.kharlamov@ifi.uio.no}
\author{Yuzhong Qu}
\affiliation{%
  \institution{National Key Laboratory for Novel Software Technology, Nanjing University, China}
}
\email{yzqu@nju.edu.cn}

\renewcommand{\shortauthors}{J. Chen et al.}

\begin{abstract}
Reusing published datasets on the Web is of great interest to researchers and developers. Their data needs may be met by submitting queries to a dataset search engine to retrieve relevant datasets.
In this ongoing work towards developing a more usable dataset search engine, we characterize real data needs by annotating the semantics of 1,947~queries using a novel fine-grained scheme, to provide implications for enhancing dataset search. Based on the findings, we present a query-centered framework for dataset search, and explore the implementation of snippet generation and evaluate it with a preliminary user study.
\end{abstract}

%
%



\maketitle

\section{Introduction}

Reusing existing datasets helps in improving productivity and reducing cost for application developers. 
In order to support convenient search of datasets that match a developer's \emph{data needs}, Google Dataset Search and other \emph{dataset search engines} have recently emerged. However, there is much room for improving their usability. 
Existing efforts mostly focus on metadata management~\cite{google}, result filtering~\cite{DBLP:conf/semco/KunzeA13}, and dataset browsing~\cite{DBLP:conf/semweb/PietrigaGADCGM18}.
At the same time little attention has been given to the queries in dataset search, which may differ considerably from general Web search queries.

\begin{figure}[t!]
  \centering
  \includegraphics[width=1\linewidth]{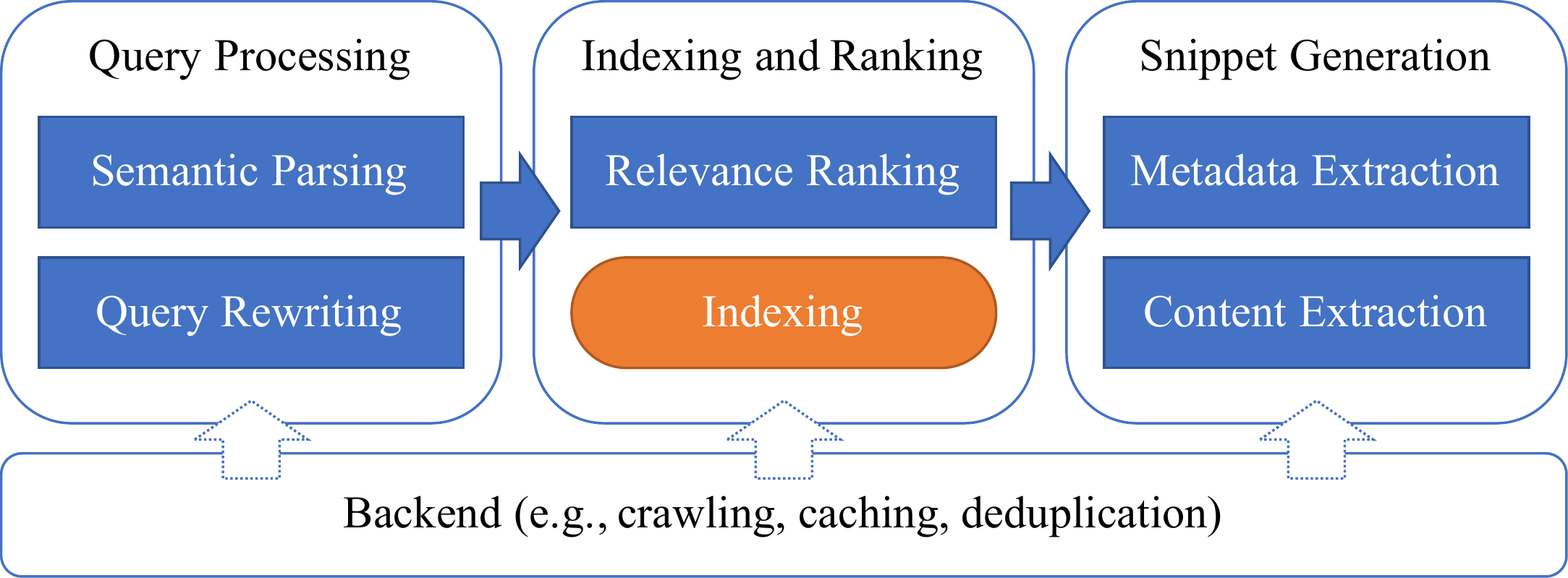}
  \caption{Proposed framework for dataset search.}
  \label{fig:framework}
\end{figure}

In this ongoing work towards developing a more usable dataset search engine, we first analyze real data needs collected from user-submitted posts on a variety of websites.
We reveal the complex constitution of a typical dataset search query, where various elements of both metadata and data content can be mentioned. 
This brings us to the idea that as a specialization of information retrieval~\cite{DBLP:conf/semco/KunzeA13}, dataset search poses unique challenges.

Based on our empirical findings, 
in Figure~\ref{fig:framework} we present a new \emph{query-centered framework} for dataset search that has four main components. (i)~\emph{Query Processing} component understands how a data need is formulated as an input query and pre-processes the query in order to achieve best query results. The query is interpreted by semantic parsing, and will be rewritten (e.g.,~relaxed) when needed (e.g.,~to avoid empty results).
(ii)~\emph{Indexing and Ranking} component determines which datasets are relevant to the query. We compute relevance and construct indexes for fast computation. (iii)~\emph{Snippet Generation} component explains how a retrieved dataset is relevant to the query and reflects the underlying data need. A query-biased snippet is extracted from its metadata and/or its content. 
(iv)~All these components are supported by the \textit{backend module}, which crawls, caches, and cleans datasets.

\begin{table*}[!t]
  \caption{Query Annotation Scheme and Its Distribution in Dataset Search Queries}
  \label{tab:scheme}
  \begin{tabular}{llrp{11cm}}
    \toprule
    \multicolumn{2}{l}{Category} & \% of Queries & Example Query \\
    \midrule
    \multirow{8}{*}{Metadata} & Name & $3.54\%$ & \emph{HUST-ASL} Dataset \\
    & Domain/Topic & $94.45\%$ & \emph{weather} dataset with solar radiance and solar energy production \\
    & Data Format & $16.23\%$ & \emph{jpg images} for all unicode characters \\
    & Language & $3.90\%$ & annotated moive review dataset in \emph{German} \\
    & Accessibility & $7.40\%$ & \emph{open source} handwritten English alphabets dataset \\
    & Provenance & $0.21\%$ & \emph{FDA datasets} about medicine name and the result has adverse events \\
    & Statistics & $2.98\%$ & dataset contains at least \emph{1000} examples of opinion articles \\
    & Overall & $96.05\%$ & \\
    \hline
    \multirow{7}{*}{Content} & Concept & $50.59\%$ & dataset about people, include \emph{gender}, \emph{ethnicity}, \emph{name} \\
    & Geospatial & $19.21\%$ & judicial decisions in \emph{France} \\
    & Other Entities & $0.41\%$ & datasets with nutrition data for many commercial food products (i.e., \emph{Lucky Charms}, \emph{Monster Energy}, \emph{Nutella}, etc.) \\
    & Temporal & $9.35\%$ & \emph{2011--2013} MoT failure rates on passenger cars \\
    & Other Numbers & $1.59\%$ & businesses that employ over \emph{1000} people in Yorkshire region \\
    & Overall & $63.79\%$ & \\
    \bottomrule
  \end{tabular}
\end{table*}

Some components in the framework have attracted research interests, such as dataset ranking~\cite{DBLP:conf/www/ToupikovUDHT09}.
Others are still rather ad hoc; they lack rigorous solutions or their performance has not been rigorously evaluated. In this paper, we report our progress in two aspects: \emph{characterizing data needs} (Section~\ref{sect:needs}) and \emph{generating dataset snippets} (Section~\ref{sect:snippet}). Our contribution is summarized as follows.
\begin{itemize}
    \item We collect real data needs from diverse sources, including user-submitted posts from online communities and data requests submitted to a national open data portal. We derive 1,947~dataset search queries and semantically annotate them using a novel fine-grained scheme. Our analysis provides implications for enhancing dataset search.
    \item For RDF data, in contrast to query-independent illustrative dataset snippet~\cite{DBLP:conf/wsdm/ChengJDXQ17}, we extract a query-biased snippet from the content of a dataset by adapting a state-of-the-art algorithm for the group Steiner tree problem~\cite{DBLP:conf/sigmod/LiQYM16}. We conduct a user study to evaluate the usefulness of these two types of snippets, and also quantitatively analyze their features.
\end{itemize}

\section{Characterization of Data Needs}\label{sect:needs}

Comparing to general Web search,
the knowledge about dataset search is rather limited. Therefore, we collect and analyze real data needs to provide empirical findings and implications for enhancing dataset search. We will now discuss this in details.

\subsection{Collection and Preprocessing}

\textbf{Collection.}
We collected descriptions of real data needs from four sources, including three online communities, namely Stack Overflow, Open Data Stack Exchange, and Reddit, and one national open data portal, namely data.gov.uk. For each online community, we leveraged its search function to retrieve posts with the query ``looking for dataset''. 
Note that since Reddit covers a wide range of topics, our search was performed within its r/datasets community. From the search results, we manually identified 50~top-ranked posts where a clear data need was described, e.g.,
\begin{quote}
    \textit{I am looking for datasets that lists the location of accidents or traffic (latitude and longitude) with date and time in many countries. I found datasets for USA and UK, now looking for datasets for other countries. Any type of road accident would be great.}\footnote{https://opendata.stackexchange.com/questions/11146/dataset-for-road-accidents-or-traffic}
\end{quote}
\noindent For data.gov.uk, we reused 50~user-submitted data requests which were sampled and published by the authors of~\cite{DBLP:conf/www/KacprzakKTS18}. To summarize, a total of 200~descriptions of data needs\footnote{http://ws.nju.edu.cn/datasetsearch/query-cikm2019/dataneeds.txt} were collected.

\textbf{Preprocessing.}
The description of a data need may be long and noisy. To improve the accuracy of analysis, we recruited ten human experts to summarize each description and remove irrelevant information. Specifically, each description was independently presented to every human expert. The expert summarized the data need by reformulating it as a concise free-form \emph{dataset search query} containing 1--20~words. 
For the 150~posts collected from online communities, we received $1,500$~reformulated queries from ten experts; an example such query is:
\begin{quote}
    \textit{location of accidents or traffic with date and time in many countries \,.}
\end{quote}
\noindent 
From the resulting set we removed 2 queries
that were arguably meaningless.
Moreover, the 50~data requests submitted to data.gov.uk had been processed in the same way by crowd workers~\cite{DBLP:conf/www/KacprzakKTS18}, producing 449~queries. To summarize, a total of $1,947$~dataset search queries\footnote{http://ws.nju.edu.cn/datasetsearch/query-cikm2019/queries.txt} were derived from the collected descriptions of data needs.

\subsection{Analysis and Implications}\label{sect:analysis}

\textbf{Annotation Scheme.}
We propose a fine-grained scheme in Table~\ref{tab:scheme} to annotate dataset search queries with their semantics. 
The scheme allows to distinguish between mentions of metadata and of data content. The latter is further divided into schema-level elements (i.e.,~concepts), instance-level elements (i.e.,~geospatial or other entities), and data values (i.e.,~temporal or other numbers). Using the proposed scheme, all the 1,947~queries have been annotated manually by two human experts. The results are accessible online\footnote{http://ws.nju.edu.cn/datasetsearch/query-cikm2019/annotations.txt}.

\textbf{Results.}
Among the 1,947 queries we collected, a query in average contains 9.22~words. More than half~($58.60\%$) of the queries contain 5-11~words. Three types of queries are identified: phrases~($63.33\%$), keywords~($31.38\%$), and sentences~($5.29\%$).

As summarized in Table~\ref{tab:scheme}, queries usually mention some metadata~(96.05\%), especially the domain/topic of interest~($94.45\%$). Data format and accessibility also occupy notable proportions~($>5\%$). On the other hand, most queries~($63.79\%$) mention the data content, mainly some schema-level concepts~($50.59\%$), followed by instance-level geospatial entities~($19.21\%$) which appear more often in the data requests submitted to data.gov.uk~($46.55\%$) but less often in other queries~($11.01\%$). Besides, temporal information is not neglectable~(9.35\%). Moreover, $60.61\%$~of the queries mention both metadata and data content.

\textbf{Implications.}
First, it would be insufficient to only take metadata into account in indexing, ranking, and result presentation like Google Dataset Search~\cite{google}. Data content should also be considered.
Second, the constitution of a dataset search query is complex, requiring novel semantic parsing techniques to process, understand, and finally improve the accuracy of search.
Third, data needs collected from different sources exhibit different features. The results of analyzing a single type of data needs (e.g.,~those submitted to national open data portals~\cite{DBLP:conf/www/KacprzakKTS18,DBLP:journals/ws/KacprzakKIBTS19}) may not be generalizable.
\section{Snippet Generation}\label{sect:snippet}

The above analysis reveals that dataset search queries mention the elements of both metadata and data content. However, current dataset search engines only present metadata about each dataset in search results pages. 
In order to complement this metadata and help the users to quickly identify relevant datasets,
we propose to extract a snippet from data content. We now present our two such methods and report preliminary evaluation results.

\subsection{Methods}
We currently focus on graph data, or more precisely RDF datasets. The content of an RDF dataset is a directed graph where nodes and edges are associated with meaningful labels.

\textbf{Query-biased Snippet.}
In the first method, a dataset search query is treated as a set of keywords after removing stop words. Each keyword is mapped to a set of nodes in the graph-structured data content. We intend to generate a query-biased snippet by extracting a subgraph that not only fully covers the query but also reflects the relationships between query keywords. We formulate the problem as finding an optimal connected subgraph that spans at least one mapped node from each query keyword. Optimality is defined by the minimization of total edge weights. We follow~\cite{DBLP:conf/icde/DingYWQZL07} to weight edges. This gives rise to an instance of the group Steiner tree problem (GST), and we implement a state-of-the-art algorithm~\cite{DBLP:conf/sigmod/LiQYM16} for solving this NP-hard problem.

\begin{table}[t!]
  \caption{Human-rated Usefulness of Snippets (1--5)}
  \label{tab:userstudy}
  \begin{tabular}{lr}
    \toprule
    & Mean $\pm$ Standard Deviation \\
    \midrule
    Query-biased Snippets & $1.91 \pm 1.22$ \\
    Illustrative Snippets & $3.04 \pm 1.23$ \\
    \hline
    $t$-test: \quad $p=0.0127$ & \\
    \bottomrule
  \end{tabular}
\end{table}

\textbf{Illustrative Snippet.}
Our second method we implement is to generate an illustrative snippet~\cite{DBLP:conf/wsdm/ChengJDXQ17} which is query-independent. It is an optimal size-constrained connected subgraph extracted from data content. Optimality is defined by a linear combination of (i)~covering the most frequent schema-level elements that appear in the data content, i.e.,~classes and properties in the RDF schema, and (ii)~covering the most central instance-level elements, i.e.,~entities. We implement the approximation algorithm presented in~\cite{DBLP:conf/wsdm/ChengJDXQ17} for solving this NP-hard problem.

\subsection{Preliminary Results}

\textbf{Design of User Study.}
We recruited 15~researchers/developers to participate in a user study. We crawled 311~RDF datasets from DataHub. 
Each participant was assigned 5~datasets and had access to their metadata and schema-level elements. For each dataset, the participant was asked to describe a data need that could be fulfilled by the dataset, by formulating a query where the keywords could be fully covered by the dataset.
Then, a query-biased snippet was computed online and compared with the precomputed illustrative snippet. The size of the former was automatically determined by the algorithm, whereas the latter was constrained to have at most 20~edges (i.e.,~RDF triples). Both snippets were blindly visualized as two node-link diagrams. The participant rated, on a scale of 1--5, the usefulness of each snippet in supporting relevance judgment.

\begin{figure}[t!]
    \centering
    \includegraphics[width = 0.4\textwidth]{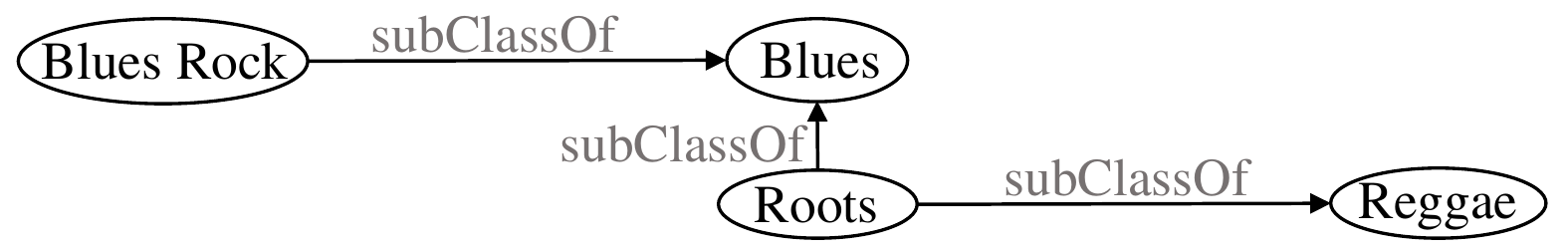}
    \caption{A query-biased snippet for the dataset \emph{Learned Genre Ontology Intl} w.r.t. the query \emph{blues rock reggae}. }
    \label{snippet}
\end{figure}

\textbf{Results.}
As summarized in Table~\ref{tab:userstudy}, illustrative snippets significantly outperform query-biased snippets~($p < 0.05$). However, they are both not very satisfying, leaving room for improvement.

Post-experiment interviews partially explained the above results.
For query-biased snippets, a snippet like the one illustrated in Figure~\ref{snippet} could cover all the query keywords, which was appreciated by the participants. However, most participants complained that such a snippet contained very limited information, due to the minimality of group Steiner trees. Indeed, in average, this kind of snippet only covered $6\%$~of the schema-level elements after weighting\footnote{Classes and properties are weighted by their frequencies in data content.}.

In contrast, for illustrative snippets, most participants confirmed their richness and the diversity of information they contain. Such snippets in average covered $75\%$~of the (weighted) classes and properties that appeared in the data content. Thus, the illustrations (or exemplifications) of schema-level concepts
that form illustrative snippets were beneficial for user comprehension of the dataset. On the other hand, illustrative snippets often failed to match the keywords appearing in the query. This kind of weak relevance to the query was criticized by the participants.

Therefore, we conclude that the two types of snippets showed complementary features. A promising direction for future work would be to combine their advantages.
\section{Related Work}\label{sect:rw}

\textbf{Dataset Search.}
Google Dataset Search~\cite{google} supports keyword search and presents the metadata for each retrieved dataset. A query is processed using the same methodology as a Web search engine, but is only matched with the metadata of a dataset. Other prototype systems~\cite{DBLP:conf/semco/KunzeA13,DBLP:conf/semweb/PietrigaGADCGM18} support faceted search.
All these systems 
exploit \emph{metadata} but ignore the \emph{content} of a dataset. By comparison, our proposed framework of dataset search is centered around the relationship between the query and the content of a retrieved dataset. We highlight semantic query parsing and query-biased snippet generation in order to understand and reflect the data need underlying a query, respectively.

\textbf{Snippet Generation.}
The utility of metadata in relevance judgment is limited because the information it spans is almost orthogonal to the content of a dataset where elements are often mentioned in a query. A promising way to complement this approach is to also present a snippet generated from the data content. IlluSnip~\cite{DBLP:conf/wsdm/ChengJDXQ17} generates an \emph{extractive} snippet by selecting a small illustrative subset of data. HIEDS~\cite{DBLP:conf/ijcai/ChengJQ16} generates an \emph{abstractive} snippet by producing a hierarchical grouping of the data content. However, none of these static snippets can precisely explain how the dataset is relevant to the query. We are among the first who explore query-biased snippet generation for datasets. Our preliminary results suggest combining the coverage of the data content~\cite{DBLP:conf/wsdm/ChengJDXQ17} with the relevance to the query~\cite{DBLP:conf/sigmod/LiQYM16} for future work.

\textbf{Query Analysis.}
Currently there is no public access to the query logs of commercial dataset search engines.
Research efforts are restricted to the queries submitted to national open data portals. In~\cite{DBLP:conf/www/KacprzakKTS18}, data requests submitted to data.gov.uk are transformed into queries by crowd workers. In~\cite{DBLP:journals/ws/KacprzakKIBTS19}, queries are extracted from the query logs of four national open data portals. The analysis in~\cite{DBLP:conf/www/KacprzakKTS18,DBLP:journals/ws/KacprzakKIBTS19}
mainly considers query length and query annotation. Their annotation scheme only includes geospatial, temporal, file/data type, numbers, and abbreviations. Their \emph{lexical} annotation is carried out automatically based on a predefined keyword mapping. They show that dataset search queries differ from general Web search queries in their length, topic, and structure. By comparison, our extended annotation scheme is more fine-grained, focusing on the \emph{semantics} of a query.
Besides, the data needs we collect are more diverse, including data requests submitted to a national open data portal as well as user-submitted posts on three online communities. Therefore, our results may exhibit better generalizability.
\section{Conclusion and Future Work}\label{sect:concl}

On the way of developing a more usable dataset search engine, we summarize our findings presented in this paper as follows.
\begin{itemize}
    \item Real data needs mention both metadata and data content.
    \item The constitution of a dataset search query is complex.
    \item Snippet generation for dataset search should combine query relevance with schema coverage.
\end{itemize}
\noindent These empirical findings and implications support our idea of designing a query-centered framework for dataset search. Extending the methods that have been implemented and priliminarily evaluated in this work, we have identified the following further steps for future development of our system.

First of all, dataset search requires specialized query processing. The analyzed data needs have exhibited some query patterns that are captured by our annotation scheme. It inspires us to formulate automated query parsing as a sequence labeling task and we plan to solve it using supervised learning techniques. To this end, a large set of data needs or dataset search queries should be labeled as training data. This in turn may reveal new patterns and requires extending our annotation scheme, until convergence.

Query processing is tightly coupled with indexing, ranking, and snippet generation. As a query may mention both the metadata and the content of a target dataset, it would be desirable to consider both of them in retrieval models and snippets. Metadata is relatively easy to handle, as it can be viewed as a semi-structured document and hence existing Web search methods may apply. By contrast, indexing, ranking, and summarizing data content pose new challenges related to both effectiveness and efficiency, as a dataset is much larger than a webpage.

With all these components being implemented, we plan to assemble a prototype of a new dataset search engine, and conduct user study to evaluate its end-to-end performance. We will start with RDF datasets, and progressively extend to other formats.

\begin{acks}
This work was funded partially by the NSFC under Grant~61772264, and partially by the SIRIUS Centre, Norwegian Research Council project number 237898. Gong Cheng was supported by the Six Talent Peaks Program of Jiangsu Province under Grant RJFW-011.
\end{acks}

\bibliographystyle{ACM-Reference-Format}
\bibliography{acmart.bib}

\end{document}